\documentclass[prl,twocolumn]{revtex4}
\usepackage{graphicx}

\begin{document}

\title{Experimental Quantum Cloning of Single Photons}
\author{Ant{\'\i}a Lamas-Linares$^{1}$, Christoph Simon$^1$, John C. Howell$^{1}$ and Dik
Bouwmeester$^{1,2}$}
\address{$^1$Centre for Quantum Computation, University of
Oxford, Parks Road, Oxford OX1 3PU, UK.\\ $^2$Department of Physics, University of California at
Santa Barbara, CA93106}

\date{28 March 2002}


\begin{abstract}
Although perfect copying of unknown quantum systems is forbidden
by the laws of quantum mechanics, approximate cloning is possible.
A natural way of realizing quantum cloning of photons is by
stimulated emission. In this context the fundamental quantum limit
to the quality of the clones is imposed by the unavoidable
presence of spontaneous emission. In our experiment a single input
photon stimulates the emission of additional photons from a source
based on parametric down-conversion. This leads to the production
of quantum clones with near optimal fidelity. We also demonstrate
universality of the copying procedure by showing that the same
fidelity is achieved for arbitrary input states.
\end{abstract}

\maketitle


No device is capable of producing perfect copies of an unknown quantum system. This statement, known
as the ``no-cloning theorem'' \cite{wootters:82,dieks:82}, is a direct consequence of the linearity
of quantum mechanics, and constitutes one of the most significant differences between classical and
quantum information. The impossibility of copying quantum information without errors is at the heart
of the security of quantum cryptography \cite{gisin:02}. If one could perfectly copy arbitrary
quantum states, this would make it possible to exactly determine the state of an individual quantum
system, which - in combination with quantum entanglement - would even lead to superluminal
communication \cite{herbert:82}. Thus the no-cloning principle also ensures the peaceful coexistence
of quantum mechanics and special relativity.

Given that perfect cloning is impossible, it is natural to ask how
well one can clone. This question was first addressed in
\cite{buzek:96}, and initiated a large amount of theoretical work.
In particular, bounds on the maximum possible fidelity of the
clones produced by universal cloning machines were derived
\cite{bruss:98}. A universal cloning machine produces copies of
equal quality for all possible input states. Following the work of
\cite{buzek:96}, quantum cloning was discussed mainly in the
language of quantum computing, where its realization was envisaged
in the form of a certain quantum logical network, consisting of a
sequence of elementary quantum gates. An implementation of the
cloning network based on NMR has recently been reported
\cite{cummins:01}, but uses ensemble techniques and thus does not
constitute true cloning of individual quantum systems. In another
experiment the polarization degree of freedom of a single photon
was approximately copied onto an external degree of freedom of the
same photon \cite{huang:01}. Although formally this is a
realization of a quantum cloning network, only a single particle
is involved in the whole process.

One might look for more natural ways of realizing quantum cloning.
In the first papers on the topic a connection to the process of
stimulated emission was made and it was suggested that stimulated
emission might allow perfect copying \cite{herbert:82}. It was
subsequently pointed out \cite{mandel:83,milonni:82} that perfect
cloning is frustrated by spontaneous emission. Recently it was
proposed \cite{simon:00} that optimal quantum cloning, where the
quality of the copies saturates the fundamental quantum bounds,
could be realized for photons using stimulated emission in
parametric down-conversion. First indications of the effect were
reported in \cite{demartini:00}, but neither universality nor
optimality were demonstrated. We present a demonstration of
universal cloning for individual quantum systems, realizing the
proposal of~\cite{simon:00} and achieving a quality of the clones
that is close to optimal.

Universal cloning by stimulated emission proceeds by sending a
single input photon into an amplifying medium capable of
spontaneously emitting photons of any polarization with equal
probability. This rotational invariance of the medium ensures the
universality of the cloning procedure \cite{simon:00}. As a result
of stimulated emission, the medium is more likely to emit an
additional photon of the same polarization as the input photon
than to spontaneously emit a photon of the orthogonal
polarization. The probabilities for stimulated and spontaneous
emission are always proportional, making it impossible to suppress
spontaneous emission without also affecting the stimulated
process. Thus, it is spontaneous emission that limits the
achievable quality of the quantum cloning and ensures that the
no-cloning theorem is not
violated~\cite{simon:00,mandel:83,milonni:82}.

\begin{figure}
\center
\includegraphics[width=\columnwidth]{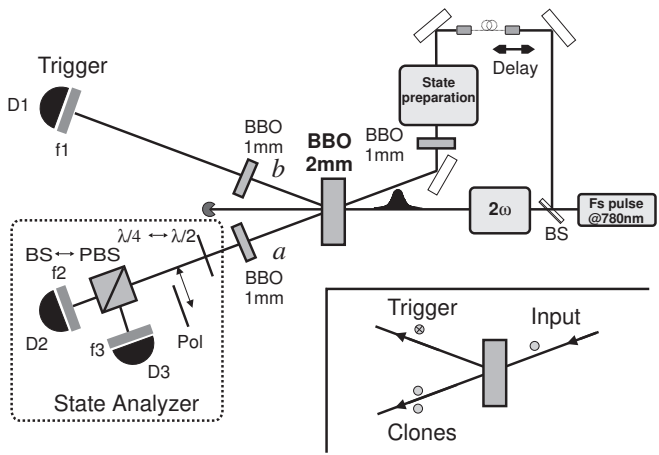}
\caption{Setup for cloning by stimulated emission. A Ti:Sapphire
laser produces light pulses of 120 fs duration, centered at a
wavelength of 780 nm. A tiny part of each pulse is split off at
the beam splitter BS and then attenuated below the single-photon
level, thus probabilistically preparing the input photon. Its
polarization state can be adjusted at will. The major part of
every pulse from the laser is frequency doubled and used to pump
the non-linear crystal (BBO 2mm), where photon pairs entangled in
polarization are created into the modes $a$ and $b$. A delay line
containing a single-mode optical fibre facilitates superimposing
the input photon and the $a$ photon produced in the crystal. For
perfect overlap, the two photons in mode $a$ after the crystal are
indistinguishable and both optimal clones of the input photon.
Their polarization is analyzed using waveplates, a polarizer and a
polarizing beam splitter (PBS) in front of detectors D2 and D3.
The photon in mode $b$ serves as a trigger, indicating that
parametric down-conversion has occurred, cf. the text. The
interference filters f1, f2, f3 help to increase the overlap
between input and down-conversion photons. The three auxiliary
crystals (BBO 1mm) compensate for birefringence in the non-linear
crystal. The inset visualizes the cloning process. Note that both
clones are in the same mode.} \label{fig:setup}
\end{figure}

The principle of our experiment is illustrated in
Fig.~\ref{fig:setup}. A strong pump light pulse propagates through
a non-linear crystal, where, with low probability, photons from
the pump pulse can split into two photons of lower frequency
(parametric down-conversion). Under suitable conditions and for
certain specific directions of emission the two created photons
are entangled in polarization \cite{kwiat:95}. The situation can
be described by a simplified interaction Hamiltonian
\begin{equation}
H=\kappa (a^{\dagger}_v b^{\dagger}_h - a^{\dagger}_h
b^{\dagger}_v) + h.c., \label{Ham}
\end{equation}
where $\kappa$ is a coupling constant, and $a^{\dagger}$ and
$b^{\dagger}$ are creation operators for photons in the spatial
modes corresponding to two different directions of emission
(Fig.~\ref{fig:setup}). The subscripts $v$ and $h$ refer to
vertical and horizontal polarization. The Hamiltonian can be shown
to be invariant under joint identical polarization transformations
in modes $a$ and $b$, ensuring that the cloning will be equally
good in every polarization basis.

The input photon arrives in mode $a$ passing through the
non-linear crystal (Fig. 1). Because of the rotational invariance
of the Hamiltonian, it is sufficient to consider one particular
initial polarization state, for example $a^{\dagger}_v
|0\rangle=|1,0\rangle_a$, where we have introduced the notation
$|k,l\rangle_a$ for a state containing $k$ vertically and $l$
horizontally polarized photons in mode $a$. Its time evolution is
obtained by applying the operator $e^{-iHt}$. For small values of
$\kappa t$, corresponding to the experimental situation, this can
be expanded into a Taylor series. The zeroth order term
corresponds to the case where no additional photons are produced.
This emphasizes that our cloning machine has a probabilistic
aspect, sometimes it will just output the input photon. The first
order term leads to the following (unnormalized) three-photon
state
\begin{eqnarray}
&-i \kappa t (a^{\dagger}_v b^{\dagger}_h - a^{\dagger}_h b^{\dagger}_v) a^{\dagger}_v |0\rangle&
\nonumber\\&= -i \kappa t ( \sqrt{2} |2,0\rangle_a |0,1\rangle_b
- |1,1\rangle_a |1,0\rangle_b).&
\label{state}
\end{eqnarray}
Recall that $|2,0\rangle_a |0,1\rangle_b$ is the (normalized)
state with 2 photons in mode $a_v$ and one photon in mode $b_h$,
while $|1,1\rangle_a |1,0\rangle_b$ has one photon each in modes
$a_v$, $a_h$ and $b_v$. Note the factor $\sqrt{2}$, which shows
that the additional emitted photon in mode $a$ is more likely by a
factor of 2 to be of the same polarization as the input photon
than of the orthogonal polarization. In this way the information
about the input photon polarization is imprinted on the
down-converted photon.

The two photons in mode $a$ are the clones. Note that in the
present ideal case the input photon and the additional photon
created in the process have identical space-time wave functions
and are thus completely indistinguishable from each other.
Therefore the two photons are both {\it approximate} copies of the
input photon with the same fidelity. Operationally, the fidelity
is defined by picking one of the two photons in mode $a$ and
determining with which probability its polarization is identical
to that of the input photon. Inspection of the output state
Eq.~\ref{state} shows that with a probability of 2/3 both photons
are vertically polarized, i.e. they are perfect clones, while with
a probability of 1/3 the photons have opposite polarization, so
that in this case the probability of picking a vertical photon is
just 1/2. Therefore the overall fidelity of the clones is given by
\begin{equation}
F=\frac{2}{3}\times 1 + \frac{1}{3} \times
\frac{1}{2}=\frac{5}{6}, \label{56}
\end{equation}
which has been shown to be the optimal achievable fidelity for the
universal cloning of a single photon \cite{bruss:98}. Note that
because of the rotational invariance of the Hamiltonian
Eq.~\ref{Ham} every other input polarization is copied with the
same fidelity.

The stimulation effect only occurs when there is overlap between
the incoming photon and the photon produced by the source. In our
experiment we use photons created in short pulses. By changing the
relative delay between the input photon and the photon created in
the down-conversion process, we can continuously vary the degree
of distinguishability. Suppose that the state of the incoming
photon, $\tilde{a}^{\dagger}_v|0\rangle$, does not overlap with
the down-conversion mode $a$. Then the same calculation as above
leads to a three-photon state
\begin{eqnarray}
&-i\kappa t (a^{\dagger}_v b^{\dagger}_h - a^{\dagger}_h b^{\dagger}_v) \tilde{a}^{\dagger}_v
|0\rangle&\nonumber\\&=
-i\kappa t (|1,0\rangle_a |1,0\rangle_{\tilde{a}} |0,1\rangle_b - |0,1\rangle_a |1,0\rangle_{\tilde{a}} |1,0\rangle_b),&
\label{state2}
\end{eqnarray}
If $\tilde{a}$ differs from $a$ only by a time delay that is small
compared to the time resolution of the detectors (which is of the
order of 1 ns), then they are practically, though not
fundamentally, indistinguishable. In this case the state in
Eq.~\ref{state2} will be experimentally indistinguishable from the
state $-i\kappa t (|2,0\rangle_a |0,1\rangle_b - |1,1\rangle_a
|1,0\rangle_b)$. Note that there is an important distinction with
respect to Eq.~\ref{state}, namely the factor $\sqrt{2}$ in the
first term has disappeared, which means that the additional
emitted photon is now equally likely to be vertically or
horizontally polarized. There is no stimulation effect.

So far, the third photon that is produced into mode $b$ has played
no role in our discussion. However, it serves an important purpose
in the experiment as a trigger. As the down-conversion photons are
created in pairs, the detection of the photon in mode $b$ means
that a clone has indeed been produced in mode $a$. For our
experimental setup, the mere detection of two photons in mode $a$
doesn't ensure that cloning has indeed occurred, because both
photons could have been contained in the input pulse. Because the
input pulse has an average photon number of only 0.05 and the
down-conversion process occurs only with a probability of the
order of 1/1000, total photon numbers larger than 3 are
exceedingly unlikely. The possible presence of more than one
photon in the input pulse leads to a slight overestimation of the
cloning fidelity (by about 0.003). However, this effect is
negligible compared to the experimental and statistical errors. It
is worth noting that, as a consequence of the anti-correlation in
polarization between the photons in modes $a$ and $b$, the photon
in mode $b$ is actually an optimal anti-clone of the input photon
\cite{buzek:99,simon:00}. Even if the phase between the two terms
in the Hamiltonian Eq.~\ref{Ham} is not fixed, such that the
entanglement between modes $a$ and $b$ is reduced, the cloning
procedure will still be universal and work with optimal fidelity,
as long as the source emits photons of any polarization with equal
probability. However, the quality of the anti-clones will steadily
decrease as the quantum correlations are lost.

In the experiment, the polarization of the photons in spatial mode
$a$ is analyzed triggered by the detection of a photon in mode
$b$, while varying the overlap between the input photon and the
photon created in the crystal. The polarization analysis is
performed in the following way. For linear polarizations a
$\lambda/2$ waveplate is used to select the measuring basis; a
polarizing beam splitter (PBS) is used to measure the events in
which the two photons in mode $a$ have different polarizations
($N(1,1)$), while a polarizer followed by an ordinary beam
splitter (BS) is used to probabilistically detect the presence of
two identical photons in mode $a$ ($N(2,0)$). For the case of
circular polarization, a $\lambda/4$ plate is used to convert
circular to linear polarization and subsequently the same method
is used. In practice, the PBS is effectively changed into a BS by
introducing an additional $\lambda/4$ to introduce minimum changes
to the experimental setup.

According to our above discussion and comparing Eqs. \ref{state}
and \ref{state2}, an enhancement of the rate $N(2,0)$ of events
where both photons have the same polarization is expected, as soon
as the input photon and the produced photon overlap. On the other
hand, for the rate $N(1,1)$ of detections where the two photons
have orthogonal polarizations, there should be no enhancement
(because the amplitude is always $i \kappa t$). Moreover, the
stimulation effect should be equally strong for all incoming
polarizations.

\begin{figure}
\center
\includegraphics[width=\columnwidth]{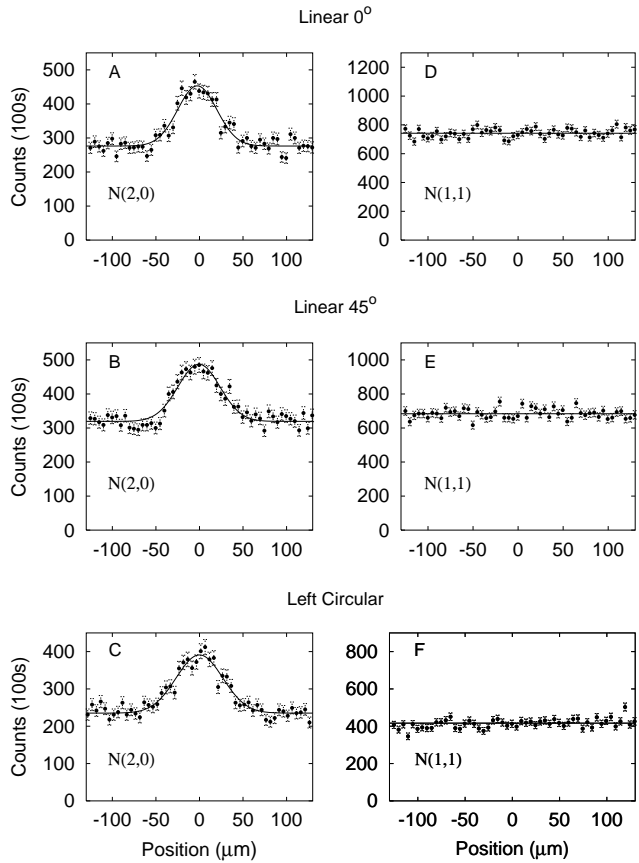}
\caption{Panels (A), (B) and (C) show the number $N(2,0)$ of
detections where both photons in mode $a$ have the polarization of
the input photon. Input polarizations were linear vertical, linear
at $45^o$, and circular left-handed respectively.  $N(2,0)$ is
plotted versus the relative distance between input and produced
photon. As expected, there is a marked increase in the overlap
region. In the ideal case of perfect overlap, the increase would
be by a factor of two. As required for universal cloning, the
enhancement is similar for each input state. The polarization
states chosen belong to three complementary bases, corresponding
to the $x$, $y$ and $z$ directions for spin. Intermediate initial
polarizations give similar results. Panels (D), (E) and (F) show
$N(1,1)$, the number of detections where the two photons have
opposite polarization, for the same three input polarizations. As
expected, $N(1,1)$ does not show any enhancement in the overlap
region. The variation of the base rates for the different inputs
is a consequence of variations in the pump power and the changes
in optical elements between the different analyzer
configurations.} \label{fig:results}
\end{figure}

These expectations are fulfilled in the experiment.
Fig.~\ref{fig:results} shows our experimental quantum cloning
results. One sees a clear increase in the $N(2,0)$ count rate in
the overlap region. This increase is observed for three
complementary input polarizations (linear $0^{\circ}$, linear
$45^{\circ}$ and circular left-handed), thus demonstrating
universality. It should be noted that far away from the overlap
region the probabilities $p(2,0)$ and $p(1,1)$ are actually the
same. This is due to the rotational invariance of the source,
which has been verified independently. The measured values for the
$N(2,0)$ and $N(1,1)$ base levels in Fig. 2 are different because
the two identically polarized photons in the $N(2,0)$ case can be
detected only probabilistically by observing coincident counts
behind a beam splitter. About half of the time, the two photons
will choose the same output port of the beam splitter and no
coincidence will be observed.

The average fidelity of the clones can be directly deduced from
Fig. 2 by taking the ratio, $R$, between the maximum and base
values in the $|2,0\rangle$ curves. The flatness of the
$|1,1\rangle$ curves demonstrates that the observed peaks are
indeed due to stimulation. From the above discussion it follows
that this is equal to the ratio between $p(2,0)$ and $p(1,1)$.
Therefore the relative probability for the two photons to have
equal polarization is $R/(R+1)$, while the probability for them to
have orthogonal polarizations is $1/(R+1)$. As a consequence, the
average fidelity of the individual clones is
\begin{equation}
F=\frac{R}{R+1} \times 1+ \frac{1}{R+1} \times
\frac{1}{2}=\frac{2R+1}{2R+2},
\end{equation}
in analogy with Eq.~(\ref{56}). The observed values of $R$ from
Fig. 2  have uncertainties of the order of $3\%$ and lead to
values for the fidelity $F$ of $0.81 \pm 0.01$, $0.80 \pm 0.01$
and $0.81 \pm 0.01$ for the three complementary polarization
directions linear vertical, linear at $45^o$ and circular
left-handed respectively. The experimental values are close to the
optimum value of $5/6=0.833$ for a universal symmetric cloning
machine. Note that strictly speaking the clones are equally good
only for perfect overlap. For imperfect overlap, one can in
principle distinguish the input photon from the photon produced by
down-conversion with a finite probability.

The absolute number of counts in Fig. 2 is determined by several
factors: the pump pulse repetition rate (80 MHz), the probability
for each input pulse to contain a photon ($5\times 10^{-2}$), the
probability of producing a down-converted pair ($10^{-3}$) and the
overall detection efficiency ($0.10$ per photon). Multiplication
of all these factors leads to the observed levels.

The limiting factor for the quality of the clones in our
experiment is the difference in (temporal) width between the input
photons and the photons produced in the down-conversion process,
leading to imperfect mode overlap. There are two reasons for this.
First, the input photon goes through several additional optical
elements which stretch the wavepacket, cf. Fig. 1. Second, the
down-conversion process intrinsically has a shorter coherence time
than the input pulse. This is largely compensated by using 5 nm
bandwidth interference filters in front of the detectors.

Another important practical point for the experiment is the
compensation for the effects of birefringence, which is achieved
by the three compensation crystals (Fig. 1). Birefringence leads
to a time delay between vertical and horizontal polarization,
which, without compensation, would considerably affect the overlap
and thus the stimulating effect for $45^o$ linear and circular
polarization. The fact that the stimulation effect  for these
polarizations is comparable to the vertical case (see
Fig.~\ref{fig:results}) indicates that the compensation is
effective.

An interesting property of universal quantum cloning machines is
that they constitute the optimal attack on certain quantum
cryptography protocols \cite{bechmann:99}. Applications of cloning
in a quantum computing context were suggested in \cite{galvao:00}.
From a more fundamental point of view, quantum cloning by
stimulated emission shows how a basic quantum information
procedure can be implemented in a natural way.

\end{document}